\documentclass[a4paper]{jpconf}
\usepackage{graphicx}

\usepackage{amssymb}

\usepackage{mathrsfs}

\usepackage{braket}

\newcommand{\D}{\mathrm{d}} 
\def\eg{{\it e.g. }} 
\def\ie{{\it i.e. }}

\newcommand{\Rh}{R_{\rm H}}

\begin{document}
\title{Horizon Quantum Mechanics: spherically symmetric and rotating sources}

\author{Andrea Giusti}
		\address{Department of Physics and Astronomy, University of 	
    	    Bologna, Via Irnerio 46, Bologna, Italy;\\ 
    	    I.N.F.N., Sezione di Bologna, IS FLAG, viale~B.~Pichat~6/2, I-40127 Bologna, Italy;\\
	    	 Arnold Sommerfeld Center, Ludwig-Maximilians-Universit\"at, 
	    	 Theresienstra{\ss}e~37, 80333 M\"unchen, Germany.}	
 		\ead{agiusti@bo.infn.it}

\begin{abstract}
In this paper we discuss some mathematical aspects of the horizon wave-function formalism, also known in the literature as horizon quantum mechanics. In particular, first we review the structure of both the global and local formalism for static spherically symmetric sources. Then, we present an extension of the global analysis for rotating black holes and we also point out some technical difficulties that arise while attempting the local analysis for non-spherically symmetric sources.
\end{abstract}

\section{Introduction}
	Black hole physics, in the classical framework, is known to have an extremely fascinating and elegant mathematical formulation in terms of Lorentzian manifold. However, such a purely classical description leads to some major physical issues such as the appearance of \textit{curvature singularities}, which is one of the main feature of black holes in general relativity \cite{Wald, Straumann}.   

	A first attempt at introducing some quantum mechanical effects in the context of Einstein's gravity was first proposed by Hawking \cite{Hawking}, whose pioneering ultimately lead to the so called \textit{semi-classical picture} of gravity. The latter basically consist in regarding black hole spacetimes as purely classical (\ie geometrical) objects on which one can study the dynamics of some quantum fields. In other words, in this framework, black holes are understood as a purely classical background stage for the quantum matter fields. Needless to say that this quantum description of gravity does not tackle the problem of curvature singularities, but rather it leads to the rise of the so called \textit{information paradox}.
	
	A way to circumvent all these issues could potentially be obtained by changing our prospective on gravity. For example, from the point of view of a particle physicist, a black hole could be understood as a gravitationally bound state confining all possible signals within its horizon. Furthermore, if we look carefully at the physical features of gravity in the high curvature regime one could convince oneself that this picture is not that different from the one of quantum chromodynamics in the strongly coupled regime. An approach which supports this viewpoint is the so called corpuscular model of quantum black holes \cite{DG-1, DG-2, PLB, PRD-1, PRD-2, PRD-3, inf, Giugno}. It is also worth remarking that this approach naturally solves the problem of singularity and the information paradox.
	
	Here, we would like to offer an alternative to the traditional semi-classical gravity by presenting the so called horizon quantum mechanics \cite{Bob, Review, Entropy} picture of gravity. Indeed, this formalism aims to a purely quantum mechanical description of trapping surfaces. In particular, the key idea is to consider the source for the black hole spacetime to be a quantum mechanical object from which we wish to infer what are the effects on the behaviour of the horizon due to the quantum nature of the source.

\section{Static spherically symmetric systems}
	Let us start then with the description of both the global and local analysis for static and spherically symmetric black holes. In particular, here we present a very short review of the main results discussed in \cite{GlobalLocal}, therefore the interested reader should refer to this paper for further details.
	
	A classical spherically symmetric spacetime can be described in terms of a line element of the form
	\begin{equation}	 
	\D s^2 = g_{tt} \, \D t^2 + g_{rr} \, \D r^2 + r^2 \, \left( \D \theta ^2 + \sin ^2 \theta \, \D \varphi ^2 \right) \, .
	\end{equation}
	The Einstein's field equation for a static source then tell us that 
	$$ g^{rr} = 1 - \frac{2 \, G_{\rm N} \, E(r)}{r} \, , \qquad E(r) = 4 \, \pi \int _0 ^r \rho ( \bar{r}) \, \bar{r}^2 \, \D \bar{r} \, , $$
	where $\rho (r)$ is the matter density of the source and $E(r)$ is the so called Misner-Sharp (or quasi-local) mass of the system\footnote{Throughout the whole paper we will set the speed of light $c=1$.}. If we now perform the study of the geodesic expansion one can easily see that the appearance of \textit{trapping surfaces} corresponds to the condition $g^{rr} = 0$ and therefore to $ r = 2 \, G_{\rm N} \, E(r)$. Furthermore, it is worth remarking that if we take the limit for $r \to \infty$, then $E(r) \to M$, where $M$ is the Arnowitt-Deser-Misner (ADM) mass of the system which enters is the definition of the Schwarzschild radius. 

	If the source is purely quantum, the quantities that define the Misner-Sharp mass and ADM mass should become quantum variables themselves. Therefore, one expects the gravitational radius will undergo the same fate, leading to a sort of quantum fuzziness of Schwarzschild (or gravitational) radius of a localised quantum source.
	
	Now, let us describe our system (source and geometry together) in terms of an Hilbert space $\mathcal{H}$ defined as
	\begin{equation}
	\mathcal{H} = \mathcal{H} _S \otimes \mathcal{H} _G
	\end{equation}
	where $\mathcal{H} _S$ and $\mathcal{H} _G$ represent the Hilbert space for the source and for the geometry, respectively. The simplest way to reproduce the connection between the geometry of the spacetime and the source, at the quantum level, is obtained by considering an entangled state
	\begin{equation}
	\mathcal{H} \ni \ket{\Psi} = \sum _{\alpha \, , \, \beta} C (E_\alpha \, , \, R_{\rm H \, \beta}) \, \ket{E_\alpha} \, \ket{R_{\rm H \, \beta}}
	\end{equation}
where the quantum states of the source correspond to the spectral decomposition of a certain hamiltonian operator $\widehat{H}$ (precisely, $\widehat{H} \otimes \mathbb{I} _G$) describing the source, while the quantum states of the geometry are labelled in terms of the spectral decomposition of a gravitational radius operator $\widehat{R} _{\rm H}$ (namely, $\mathbb{I} _S \otimes \widehat{R} _{\rm H}$). In other words, we have lifted the ADM mass and the Schwarzschild radius to operators on $\mathcal{H}$.

	However, not all states of $\mathcal{H}$ can be seen as physical states for the system within this formalism. Indeed, the only physical states are the one that allow us to gain a Schwarzschild-like relation between the spectrum of $\widehat{H}$ and the one of $\widehat{R} _{\rm H}$, \ie the quantum version of the relation between the ADM mass and the gravitational radius. The physical configurations are therefore determined by the constraint
	\begin{equation}
	\left( \widehat{H} - \frac{1}{2 \, G_{\rm N}} \, \widehat{R} _{\rm H} \right) \, \ket{\Psi} _{{\footnotesize \texttt{phys}}} = 0 \, ,
	\end{equation}
	that leads to $C (E_\alpha \, , \, R_{\rm H \, \beta}) = C (E_\alpha \, , \, 2 \, G_{\rm N} \, E_\alpha) \, \delta _{\alpha \beta}$.

	Now, if we trace out the contribution of the gravitational degrees of freedom we get
	\begin{equation}
	\ket{\Psi} _{\rm S} = \sum _\gamma C_S (E_\gamma) \, \ket{E_\gamma} \, , 
	\quad C_S (E_\gamma) \equiv C (E_\gamma \, , \, 2 \, G_{\rm N} \, E_\gamma)
	\end{equation}
	which represents the spectral decomposition of the source, whereas if we trace away the contribution of the source we get
	\begin{equation}
	\ket{\Psi} _{\rm H} = \sum _\gamma C_S (R_{\rm H \, \gamma} / 2 \, G_{\rm N} ) \, \ket{R_{\rm H \, \gamma}} \, ,
	\end{equation}
which is known as the \textbf{horizon wave-function} (HWF) and represents the fundamental ingredient of the horizon quantum mechanics.

	This object allows us to provide a \textit{probabilistic definition of black hole}. Indeed, if we assume a continuous spectrum for our observables, the HWF in the gravitational radius representation will be given by
	\begin{equation}
	\Psi _{\rm H} (R_{\rm H}) := \braket{R_{\rm H} \, | \, \Psi _{\rm H}} = \mathcal{N} _{\rm H} \, C_S (R_{\rm H \, \gamma} / 2 \, G_{\rm N} ) \, ,
	\end{equation}
where $\mathcal{N} _{\rm H}$ is just a normalisation constant. Then, we can build a \textit{probability density function} that one would detect a gravitational radius of size $R_{\rm H}$ associated with the particle in the quantum state $\ket{\Psi} _{\rm S}$, \ie
\begin{equation}
\mathcal{P} _{\rm H} (r_{\rm H}) := 4 \, \pi \, r_{\rm H} ^2 \, |\Psi _{\rm H} (R_{\rm H})| ^2 \, .
\end{equation}

	Hence, we can now define the \textit{probability density for a quantum particle to be a black hole} as the conditional probability density that the particle lies inside its own gravitational radius $R_{\rm H}$, \ie
	\begin{equation} \label{pless}
	\mathcal{P} _{<} (r < r_{\rm H}) := \mathcal{P} _{\rm S} (r < R_{\rm H}) \, \mathcal{P} _{\rm H} (R_{\rm H}) \, ,  
	\end{equation}
where 
$$\mathcal{P} _{\rm S} (r < r_{\rm H}) = 4 \, \pi \int _0 ^{R_{\rm H}} \bar{r} ^2 \, \Psi _{\rm S} (\bar{r}) \, 
{\rm d} \bar{r} \, .$$
Then, the \textit{probability for a quantum source to be a black hole} is just given by integrating (\ref{pless}) over all possible values of $R_{\rm H} \in (0, \infty)$.

	This formalism also allows for a local description of the gravitational radius, \ie for localised energy eigenmodes and correspond- ingly discrete energy quantum numbers. To see that, let us restrict our attention to what happens within a sphere of radius $r$. In this case, the role the ADM mass is now played by the Misner-Sharp mass $E (r)$, therefore the new hamiltonian operator will now be dependent on the radius, \ie $E (r) \, \mapsto \, \widehat{H} (r)$. It is easy to prove that the spectrum of $\widehat{H} (r)$ will be given by
	\begin{equation}
	\widehat{H} (r) = \sum _\alpha P _\alpha (r) \, E_\alpha \, \ket{E_\alpha} \bra{E_\alpha} \, , 
	\quad P _\alpha (r) := 4 \, \pi \int _0 ^{r} \bar{r} ^2 \, \Psi _{E _\alpha} (\bar{r}) \, {\rm d} \bar{r} \, .
	\end{equation}
Notice that $\lim _{r \to \infty} P _\alpha (r) = 1$ only for $\Psi _{E _\alpha} \in L _2 (0, \infty)$, \ie for localised energy eigenmodes. This restriction was not a requirement in the global case, since the norm of these modes never entered explicitly in that calculation.

	If we proceed as above, we can ultimately provide a local (quantum-mechanical) characterisation of trapping surfaces, specifically one finds that
	\begin{equation}
	\bra{r_{\rm H}} r_{\rm H} (r) \ket{r_{\rm H}} = r \, , \qquad  
	r_{\rm H \, \alpha} (r) := P _\alpha (r) \, R_{\rm H \, \alpha} \, ,
	\end{equation}
which is the local quantum version the condition $g^{rr} = 0$ in terms of the Misner-Sharp mass.

\section{Rotating sources: global formalism}
	The prototype for a rotating black hole, in general relativity, is given by the Kerr solution of the vacuum Einstein's field equations, \ie 
	\begin{eqnarray}
	{\rm d} s^2 &=& - \left( 1 - \frac{r_s \, r}{\Sigma} \right) \, {\rm d} t^2 + \frac{\Sigma}{\Delta} \, {\rm d} r ^2 
	+ \Sigma \, {\rm d} \theta ^2\\
	&+& \left( r^2 + a^2 + \frac{r_s \, r \, a^2}{\Sigma} \, \sin ^2 \theta \right) \sin^2 \theta \, {\rm d} \phi ^2 
	- \frac{2 \, r_s \, r \, a \, \sin^2 \theta}{\Sigma} \, {\rm d} t \, {\rm d}\phi
	\end{eqnarray}
where $M$ is the ADM mass, $r_s = 2 \, G_{\rm N} \, M$, $a = J/M$ with $J$ the angular momentum of the system, $\Sigma = r^2 + a^2 \, \cos ^2 \theta$ and $\Delta = r^2 - r_s \, r + a^2$.

For this spacetime, the horizons are given by the condition $\Delta = 0$, which in turns leads to
\begin{equation}
R_{\rm H} ^{(\pm)} = G_{\rm N} \left( M \pm \sqrt{M^2 - \frac{J^2}{M^2}} \right)
\end{equation}

As stated in the Carter-Robinson uniqueness theorem, a vacuum black hole solution in an stationary, axisymmetric and asymptotically flat spacetime belongs to the Kerr family of solutions vacuum Einstein's field equations, which in turns is parametrised by two real parameters $(M, J)$, \ie the ADM mass and the total angular momentum.

Therefore, if we want to perform an analysis in the framework of the horizon quantum mechanics \cite{HQM-Rotating}, we have to find a way to promote the two parameters $(M, J)$ to quantum mechanical observables. However, this should be done with a bit of care due to the fact that the notion of angular momentum in quantum mechanics is extremely different from the classical one.

	So, let us assume the existence of a complete set of commuting operators connected with quantum source, \ie $\{ \widehat{H}, \, \widehat{J} ^2, \, \widehat{J}_z \}$, respectively the Hamiltonian, the total angular momentum squared and the angular momentum along the axis of rotation.
	
	Then, we can just promote the two parameters $(M, J)$ to operators on $\mathcal{H}$ as follows
	\begin{equation}
	M \, \mapsto \, \widehat H
=
\sum _{a, j , m} E_{a \, j} \, \ket{a \, j \, m} \bra{a \, j \, m}
\ ,
\end{equation}
\begin{equation}
J^2 \, \mapsto \, \widehat{J} ^{2} 
=
\, m _{\rm p} ^{4} \, \sum _{a, j , m} j (j+1) \, \ket{a \, j \, m} \bra{a \, j \, m} \equiv \sum _{a, j , m} \lambda _{j} \, \ket{a \, j \, m} \bra{a \, j \, m} ,
\end{equation}
	\begin{equation}
\widehat{J} _{z} =
m _{\rm p} ^2 \,\sum _{a, j , m}  m \, \ket{a \, j \, m} \bra{a \, j \, m} \equiv \sum _{a, j , m} \xi _{m} \, \ket{a \, j \, m} \bra{a \, j \, m} \,
\ .
	\end{equation}
	where $j \in \mathbb{N} _{0} /2$, $m \in \mathbb{Z} / 2$, with $|m| \leq j$,
and $a \in \mathcal{I}$, with $\mathcal{I}$ a \textit{discrete}
set of labels that can be either finite or infinite.

	In analogy with the previous section, the quantum state for the whole system (source plus geometry) can be expressed as a triple entangled state, \ie
	\begin{equation}
	\ket{\Psi}
=
\sum_{a, j , m} \sum _{\alpha , \beta}
C(E_{a\, j}, \lambda _{j} , \xi _{m}, {\Rh} ^{(+)} _\alpha , {\Rh} ^{(-)} _\beta) \,
\ket{a \, j \, m}
\ket{\alpha}_{+}
\ket{\beta} _{-} 
	\end{equation}
	where $\ket{\alpha}_{\pm}$ are the eigenstates of $\widehat{R} _{\rm H} ^{(\pm)}$.
	
	Now, following the same line of thought as before, we can select the physical states for our system by imposing two constraints on $\ket{\Psi} \in \mathcal{H}$, namely 
\begin{equation}
\left(\widehat R_{\rm H} ^{(+)} - \widehat{\mathcal{O}} ^{+} \right)
\ket{\Psi} _{\rm phys}
=
0 \, , 
\quad
\left(\widehat R_{\rm H} ^{(-)} - \widehat{\mathcal{O}} ^{-}\right)
\ket{\Psi} _{\rm phys}
=
0 \, ,
\end{equation}
with
$$\widehat{\mathcal{O}} ^{\pm}:= \widehat H \pm \left( \widehat{H} ^{2} - \widehat{J} ^{2} \, \widehat{H} ^{-2} \right) ^{1/2} \, ,$$
provided that $\widehat{\mathcal{O}} ^{\pm}$ is positively semi-definite.
	
	Then, following the same procedure as in the previous section, we can now compute the (unnormalized) HWF for each horizon and, consequently, the black hole probability. 
	
	It is worth remarking that, due to the lack of a proper generalization of the Misner-Sharp mass for non-spherically symmetric systems, the local formalism for rotating sources is still an open problem.
	
	\section{Conclusions and outlook}
	We have been able to provide a precise mathematical formulation of both the global and local horizon quantum mechanics for static spherically symmetric quantum sources. Specifically, we have introduced the notion of probabilistic definition of black hole and we have provided a complete characterisation of trapping surfaces at the quantum level.
	
	Furthermore, we have also been able to extend the global formalism to the case of rotating sources described in terms of a metric of the Kerr family. However, a complete quantum mechanical local analysis requires a generalised notion of Misner-Sharp mass for axisymmetric system which is, however, still an open problem in classical general relativity.
	
	Despite the many works talking the implementation of the horizon quantum mechanics for classical solutions of Einstein's field equations, very few papers have been dealing with metric describing non-singular spacetimes. Therefore, it would be interesting to see an application of this formalism to the Hayward spacetime \cite{Hayward, GGH}, and its rotating counterpart (see \eg \cite{rot-hay, mio}).

\end{document}